\documentclass[a4paper,11pt]{article}
\usepackage{jcappub} 
\usepackage[T1]{fontenc}
\usepackage{gensymb}
\usepackage{aas_macros}

\title{New constraints on modelling the random magnetic field of the MW}
\author[a,1]{Marcus C. Beck,\note{Corresponding author.}}
\author[b]{Alexander M. Beck,}
\author[c]{Rainer Beck,}
\author[b,d]{Klaus Dolag,}
\author[e]{Andrew W. Strong,}
\author[a]{and Peter Nielaba}

\affiliation[a]{Department of Physics, University of Konstanz,\\Universit\"{a}tsstr. 10, D-78457 Konstanz, Germany}
\affiliation[b]{University Observatory Munich,\\Scheinerstr. 1, D-81679 Munich, Germany}
\affiliation[c]{Max Planck Institute for Radioastronomy,\\Auf dem H\"{u}gel 69, D-53121 Bonn, Germany}
\affiliation[d]{Max Planck Institute for Astrophysics,\\Karl-Schwarzschild-Str. 1, D-85741 Garching, Germany}
\affiliation[e]{Max Planck Institute for Extraterrestrial Physics,\\Giessenbachstr. 1, D-85748 Garching, Germany}

\emailAdd{marcus.beck@uni-konstanz.de}
\emailAdd{abeck@usm.uni-muenchen.de}
\emailAdd{rbeck@mpifr-bonn.mpg.de}
\emailAdd{dolag@usm.uni-muenchen.de}
\emailAdd{aws@mpe.mpg.de}
\emailAdd{peter.nielaba@uni-konstanz.de}

\arxivnumber{1409.5120}

\abstract{
  We extend the description of the isotropic and anisotropic random
  component of the small-scale magnetic field within the existing
  magnetic field model of the Milky Way from Jansson \& Farrar, by
  including random realizations of the small-scale
  component. 
  Using a magnetic-field power spectrum with Gaussian
  random fields, the NE2001 model for the thermal electrons and the
  Galactic cosmic-ray electron distribution from the current GALPROP
  model we derive full-sky maps for the total and polarized
  synchrotron intensity as well as the Faraday rotation-measure
  distribution. 
  While previous work assumed that small-scale fluctuations
  average out along the line-of-sight or which only computed ensemble
  averages of random fields, we show that these fluctuations need to be
  carefully taken into account. 
  Comparing with observational data we
  obtain not only good agreement with 408 MHz total and WMAP7 22 GHz
  polarized intensity emission maps, but also  an improved
  agreement with Galactic foreground rotation-measure maps and power
  spectra, whose amplitude and shape strongly depend on the parameters of
  the random field. 
  We demonstrate that a 
  correlation length of $\approx220$~pc ($50$~pc being a 5$\sigma$
  lower limit) is needed to match the slope of the observed power
  spectrum of Galactic foreground rotation-measure maps. Using
  multiple realizations allows us also to infer errors on individual
  observables. 
  We find that previously-used amplitudes for random
  and anisotropic random magnetic field components need to be rescaled by
  factors of $\approx{0.3}$ and $0.6$ to account for the new
  small-scale contributions.
  \linebreak\newpage\noindent
  Our model predicts a rotation measure of
  $-2.8\pm7.1$ rad/m$^2$ and $4.4\pm11.0$ rad/m$^2$ for the north and
  south Galactic poles respectively, in good agreement with observations.
  Applying our model to deflections of
  ultra-high-energy cosmic rays we infer a mean deflection of
  $\approx3.5\pm1.1$ degree for  60 EeV protons arriving from
  CenA.
}


\begin{document}
\maketitle
\flushbottom


\section{Introduction}\label{sec:intro}

Magnetic fields are an important property of most astrophysical systems including the Milky Way Galaxy.
Over the past years we have gained increasing insight into the magnetic field of our Galaxy and its structure on large and  small scales.
From radio observations (see e.g. \cite{beck96}) we can constrain the Galactic magnetic field (GMF) and the physical processes responsible for its creation.
Within galaxies in general, and thus also within our Milky Way, magnetic fields are thought to be  generated by dynamo action.
These dynamos operate on very small seed fields, which may already have been present even before the first galaxies appeared in the Universe (for reviews on cosmic magnetism see e.g. \cite{beck13b,widrow02}).
The detailed structure of the GMF is still under debate, but recent observational advancements have significantly improved our knowledge of large-scale and small-scale features of the GMF (e.g. with the PLANCK mission \cite{planck14a,planck14b,planck14c,planck14d,planck14e}).
From the extensive literature on the GMF we mention the  reviews by \cite{haverkorn14,noutsos12,heiles12}. 
The total GMF is usually split-up into several small-scale and large-scale fields.
In the literature on this topic,  differing and thus confusing notations are used for these components: the large-scale field is often called regular, uniform or coherent (see e.g. \cite{jaffe10}) and the small-scale fields are referred to as random, tangled or turbulent fields. 
Recently another type of field has been considered to complement the small-scale field: this is referred to as anisotropic random (see e.g. \cite{sokoloff98,orlando13}), ordered random (see e.g. \cite{jaffe10}) or striated field (see e.g. \cite{jf12a}).
The distribution of random field vectors is characterised by a zero mean field vector but a non-zero RMS (dispersion) value.
This holds for all random field types but the crucial difference between anisotropic and isotropic random fields is in the dispersion of the vectors for all three space dimensions. 
For anisotropic fields a direction can be found in which the dispersion is much smaller (see also Sec. 5.1.1 in \cite{sokoloff98}), 
while for isotropic random fields the dispersion is the same for all directions.
Possible candidates for the origin of such fields  are compression or shear of magnetized turbulent gas e.g. by supernova shocks or density waves in spiral arms (see e.g. \cite{haverkorn14}).
The clearest indication of anisotropic random fields comes from  radio observations of M51 \cite{fletcher11}.
Important contributions to understanding the GMF and its turbulent component have been made by \cite{gaensler11} as well as \cite{iacobelli14} who compared observational polarization gradient maps to data from idealised numerical magnetohydrodynamic (MHD) turbulence calculations.
The various components of the total magnetic field can be distinguished by their differing contributions to observables such as total synchrotron intensity (I), polarized intensity (PI) and rotation measure (RM) (see e.g. \cite{beck13b} and Tab. \ref{tab:models}).
The anisotropic random field as an additional component is constructed to contribute to total and polarized synchrotron emission,
and therefore to allow for smaller, more realistic cosmic-ray electron densities to match observations.
On the other hand it does not increase RM values, but increases its scatter. 
The goal of the introduction of an anisotropic component is to close the gap between too high RM respectively too low I,PI and therefore reach consistency between simulations and observations.
It is very important to constrain the modelling process not only with single observables (at a single frequency) but preferably at several at once.
The comparison of  synchrotron emission and RM to simulated data is  well-suited  to constrain the various magnetic field components, not only due to the fact that synchrotron emission is dependent on the perpendicular (with respect to the line-of-sight) magnetic field and RM on the parallel component, but also because of the different effects of cosmic-ray electron and thermal  electron distributions.

Furthermore, it is believed that small-scale fluctuations will change the propagation of ultra high energy cosmic rays (UHECR) and play an important role in the analysis thereof (see e.g. \cite{tinyakov05}). For this a detailed knowledge on the structure of the small-scale field is also required.

Until recently, most studies focussed on modelling the large-scale GMF (see e.g. \cite{page07}).
Recently the small-scale field component has received more  attention:
\cite{miville08} use 408 MHz total synchrotron data and 22 GHz WMAP polarized synchrotron data to study a bi-symmetric spiral structure (BSS) model of the large-scale halo field and a random component with about half the amplitude of the regular component.
In \cite{jaffe10} 408 MHz I data, WMAP 22 GHz PI data and 269 extragalactic RM sources are considered. They found support for a spiral structure of the large-scale disk field,  and random and anisotropic small-scale field strengths  five and  four times as large as the regular component respectively.
\cite{fauvet11,fauvet12} considered WMAP 5 I 22 GHz, ARCHEOPS 353 GHz and 408 MHz I data and found a modified axisymmetric logarithmic spiral (ASS) structure and an upper limit for the random small-scale amplitude of a quarter of the large-scale amplitude while the data seem to indicate no turbulent magnetic field.
\cite{sun08} found random fields with a strength of about 3~$\mu\mathrm{G}$ to give the best fit to the ASS model for the disk, and found an reversed symmetry in the halo.
In \cite{sun09} their model was used to simulate 1.4 GHz synchrotron emission at arcsec angular resolution for the forthcoming Square Kilometre Array (SKA),
 and I, PI and RM of selected patches of the sky are studied.
The random magnetic field component  was modelled to follow a Kolmogorov spectrum and no anisotropic random field was considered.
In \cite{orlando13}, besides different large scale models (ASS, ASS+RING and BSS) the random field was assumed to be exponential in height as well as in the radial direction. 
The anisotropic random field was assumed to follow the topological form of the regular field.
Recently, \cite{mertsch13} modelled small-scale structures of the 408 MHz synchrotron emission with an angular power spectrum
based on GMF models and the distribution of relativistic electrons from \cite{strong11}. 
A large-scale distribution of magnetic field and relativistic electrons an additional turbulent magnetic spectrum was introduced.
To increase the power on medium angular scales, they added synchrotron emission from isolated supernova remnants.

Recently models with different magnetic field configurations for the disk as well as for the halo have been developed.
\cite{jf12a,jf12b} combined significantly differing models for the Galactic disk and halo.
Their best-fit model has an even symmetry in the disk and an odd symmetry in the halo.
We base our modifications and advancements on their - already very sophisticated - model for the GMF, which is denoted as JF12 hereafter.

\begin{table}
\centering
\small
\begin{tabular}{|l|l|p{3.6cm}p{3.6cm}ll|}\hline
Component & Model & Large-scale features & Small-scale features & Observabl. & Ref.\\\hline
$n_\mathrm{e}$ & NE2001	& Spiral arms, molecular ring & Galactic-centre, local ISM structures, under-dense regions & RM & \cite{ne2001}\\
$n_\mathrm{CRe}$ & GALPROP & Distribution, 54\_z04LMPDS & - & I, PI & \cite{strong10}\\
$B_\mathrm{reg}$ & JF12a & Spiral arms, toroidal-, X-halo & - & I, PI, RM & \cite{jf12a}\\
$B_\mathrm{iso}$ & JF12b & Spiral arms,  halo & - & I & \cite{jf12b}\\
$B_\mathrm{aniso}$ & JF12a & $B_\mathrm{aniso}^2 = \beta B_\mathrm{reg}^2$ & - & I, PI & \cite{jf12a}\\\hline
\end{tabular}
\caption{Summary of  models  are used in this analysis of the GMF.}\label{tab:models}
\end{table}

To constrain the GMF models, we have first Faraday Rotation Measures (RM) of extragalactic sources. 
RM values are observed integrals along the line-of-sight (LOS) $\ell$
\begin{equation}
RM \propto \int_\ell n_\mathrm{e}(l) B_\parallel(l) ~\mathrm dl
\end{equation}
The  RM gives the wavelength-dependent amount of Faraday rotation of the polarization vector of the radio emission of the source with polarization angle $\Psi_0$. The polarization angle $\Psi$ is then given by $\Psi = \mathrm{RM} \cdot \lambda^2 + \Psi_0$.
By convention, RM is positive (negative) if the magnetic field is directed towards (away from) the observer.
It should to be noted that each extragalactic source may have intrinsic RMs.
Another observational constraint on the GMF is the synchrotron emission which results from relativistic electrons that gyrate around magnetic field lines and  emit linearly polarized radio radiation.
Therefore the (polarized) synchrotron emission depends on the transverse (to the LOS) magnetic field component $B_\perp$ and the density of the relativistic cosmic-ray electrons $n_\mathrm{CRe}$.
The polarized intensity is computed as
\begin{equation}
PI \propto \int_\ell n_\mathrm{CRe}(l) B_\perp^2(l) ~\mathrm dl.
\end{equation}

We use the HAMMURABI code \cite{waelkens09} to compute the above integrals and calculate full-sky mock observations of various models of the Galactic magnetic field, the cosmic-ray (CR) electron density $n_{\mathrm{CRe}}$ and the thermal electron density $n_\mathrm e$.
We compute Galactic rotation measure (RM), synchrotron maps of total intensity (I), Stokes parameters Q and U (and therefore polarized synchrotron intensity (PI) and polarization angle (PA)) and deflection maps for ultra high energy cosmic rays.

For the thermal electron density we use the  NE2001 model \cite{ne2001}.
This model is constrained by dispersion measures of pulsars, which are proportional to the modelled electron density along the LOS to the pulsar. The arrival time of the pulsed emission  is dependent on the density of free electrons.
The most prominent large-scale features of the model are spiral arms and a molecular ring component. There are only a few small-scale features included: a Galactic-centre component and local ISM structures and under-dense regions.
There are no small-scale fluctuations of $n_e$ in the NE2001 model. 
Although it has been shown by \cite{gaensler08} that the NE2001 distribution is not accurate and it was suggested to increase the vertical scale height by a factor of two, we use the original model parameters because the GMF model parameters of the model on which we base our investigation on was constrained by using the original NE2001 parameters
 (so the RM maps of JF12 could only be reproduced using the original parameters).

The model for $n_\mathrm{CRe}$ used here is identified by its GALPROP parameter set  54\_z04LMPDS and is described in detail in \cite{strong10}.
For a recent review of cosmic rays including electrons and synchrotron putting this in context see \cite{AnnRev2015}.
The CR model was computed with the GALPROP software; for a full account of the method see \cite{strong07}.\footnote{see http://galprop.stanford.edu; latest versions available  at http://sourceforge.net/projects/galprop}
Models using the same electron distribution are extensively compared with synchrotron data in \cite{orlando13}.
Briefly, this model uses a radial Galactic distribution of CR electron sources based on the distribution of pulsars, since these are a tracer of supernova remnants, which are thought to be the sources of CR.
The radial distribution is  extended to $R=15$ kpc to account for gamma-ray and synchrotron observations of the outer Galaxy as described in  \cite{orlando13}.
The propagation region has a vertical height of 4 kpc from the Galactic plane and a radius of 20 kpc. 
CR propagation is for pure diffusion without reacceleration, since this is found to give better agreement with synchrotron radiation from the Galaxy  \cite{strong11}.
The electron spectrum in this model is  consistent  with the measurement at the solar position with Fermi-LAT, in the range 10 GeV to 1 TeV. 

The synchrotron emission at 408 MHz originates from CR electrons  with energies of a few GeV in a magnetic field of around 5 $\mu$G.
The interstellar CR electron spectrum is not well known at these energies due to solar modulation  in the heliosphere, and the synchrotron production
function is very broad, so it is better to use the synchrotron spectrum itself as a guide to the effective mean index after integration over the entire electron spectrum.
The synchrotron intensity index at 408 MHz is about 0.8 (see plots and Appendix of \cite{strong11}), which corresponds to an electron index of $p=2.6$ instead of 3.
Since synchrotron brightness temperature is proportional to $B^{(p+1)/2}$, this gives a difference of 0.2 in the power of B.
This difference is negligible compared with other uncertainties, so that the use of $p=3$ is a reasonable approximation.
But in future it would be desirable to use the full synchrotron calculation from the GALPROP electron spectrum in HAMMURABI.

RM observations are a useful tool to study the GMF and its halo (see e.g. \cite{pshirkov11,mao10,mao12}).
In particular, small-scale fluctuations within the RM data (see e.g. \cite{schnitzeler10,oppermann12,oppermann14}) can be used to study the random and turbulent components of the GMF.
The model of \cite{jf12a,jf12b} is however not  able to reproduce fluctuations in the full-sky maps.
In the present work we show how to overcome this problem and present an improved formulation for the small-scale GMF.

We explain our general advances in modelling the GMF  by taking the example of the JF12 model.
A significant improvement resulting from our approach is the use of an ensemble of random field realizations.
Up to now, most studies employed only one realization and the averaging over an ensemble was done by performing line of sight integrations, which does not represent an ensemble of possible small-scale fields.
With our approach we can specify uncertainties in the observables such as the angular power spectrum of the map.
We claim that such uncertainties should always be considered in future. 
As an application, an uncertainty map of the deflection angles of UHECR can  be estimated with our method.


\section{Modelling the Galactic magnetic field}
\label{sec:modifications}
We base our investigation on the large-scale GMF model of \cite{jf12a}.
Their model contains separate components for the magnetic field patterns within the spiral disk, the toroidal halo field and the X-shaped poloidal halo field.
Each of these components are modelled in a way to ensure zero magnetic divergence.
The detailed method for obtaining both the parameters and the quality of the fits of the models are given in \cite{jansson09}.
The main difference between the method of JF12 and our approach is the way the small-scale fields are incorporated in the modelling process. 
First we describe the JF12 large-scale field model and then describe these differences in the implementations of JF12 and our approach.

We label the original field components from our base models with the index `JF12' and make use of the following variable names:
$B_\mathrm{reg}$ represents the large-scale field and $B_\mathrm{aniso}$ and $B_\mathrm{iso}$ represent the (an)isotropic small-scale fields.
The corresponding small-scale models in JF12 are: $B_\mathrm{striated,JF12}$ and $B_\mathrm{rand,JF12}$.
Unchanged field components in our modifications are still denoted as JF12.
Table \ref{tab:JF12parameters} summarizes the used parameters.
\begin{table}\centering
	\begin{tabular}{|p{1.8cm}lp{3cm}|p{1.8cm}lp{3cm}|}
	\hline
	\multicolumn{3}{|c|}{Large scale field}  & \multicolumn{3}{|c|}{Isotropic random field}\\\hline
	Disk \newline component & $b_{\{1-8\}}$  & \{0.1, 3.0, -0.9, \newline-0.8, -2.0, -4.2, \newline0.0, 2.7\} $\mu$G & 
	Disk \newline component &$b_{\{1-8\}}$   & \{10.81, 6.96, 9.59, \newline6.96, 1.96, 16.34, \newline37.29, 10.35\} $\mu$G\\
	        & $b_\mathrm{ring}$  &  0.1  $\mu$G &            &  $b_\mathrm{int}$  &  7.63  $\mu$G\\
	        & $h_\mathrm{disk}$  &  0.40 kpc &               &  $z_0^\mathrm{disk}$  &  0.61 kpc        \\      
	        & $w_\mathrm{disk}$  &  0.27 kpc &               Halo & $B_0$ & 4.68 $\mu$G       \\ 
	Toroidal & $B_\mathrm{\{n|s\}}$   & 1.4 $\mu$G | -1.1 $\mu$G		&component& $r_0$ & 10.97 kpc  \\
	halo	& $r_\mathrm{\{n|s\}}$  & 9.22 kpc | 16.7 kpc 			&&$z_0$ & 2.84 kpc \\
		& $w_\mathrm{h}$  & 0.20 kpc 			&&&  \\\cline{4-6}
		& $z_0$  & 5.3 kpc				&\multicolumn{3}{|c|}{Anisotropic random field}\\\cline{4-6}
	X halo  & $B_\mathrm X$  & 4.6 $\mu$G			&Striation	&$\gamma$&2.92 \\
		& $\Theta_\mathrm X^0$  & 49 $^\circ$		&		&$\beta$&1.36 \\
		& $r_\mathrm X^0$  & 4.8 kpc			&		&\multicolumn{2}{l|}{$B_\mathrm{aniso}^2=\beta B_\mathrm{reg, JF12}^2$} \\
		& $r_\mathrm X$  & 2.9 kpc			&&& \\
	\hline
	\end{tabular}
	\caption{Summary of the used GMF model parameters for the large and small scale fields.\label{tab:JF12parameters}}
\end{table}

\subsection{JF12 large-scale model}\label{sec:JF12large-scale}
We use coordinates that are either Cartesian ($x$, $y$, $z$) or cylindrical ($r$, $\phi$, $z$). The Galactic centre is located at the origin of both coordinate systems. The $z$-axis defines the direction towards Galactic North Pole and the $x$-axis points away from the position of the Sun which is situated at $x=-8.5$ kpc.

\subsubsection{Spiral disk}
The spiral disk is modelled  using the functional form in \cite{brown07}.
The spiral arms therein are based on the structural form of the NE2001 electron distribution model. 
\footnote{In external spiral galaxies significant decrease of the magnetic pitch angle with glalctocentric radius is found (e.g. \cite{fletcher04} found that in M31 the pitch angle decreases from -19$^\circ$ near the center to -8$^\circ$ at about 12-14 kpc).
Other successful models are constructed that account for this finding, e.g. \cite{eck11} which revisited the spiral disk field by studying the Galactic disk field in three different sectors. Thereby the inner field still has a spiral magnetic field that is aligned with the spiral arms. The outer disk is dominated by an almost purely azimuthal field.}
Eight spiral arm segments are defined by logarithmic curves which defines their dividing boundaries; they are characterized by the equation $r = r_{-x}e^{\phi / \tan(90^\circ-i)}$ with inclination angle $i=11.5^\circ$. 
Each dividing line has its own starting point $r_{-x}$ and field strength $b_i$ which defines the magnetic field strength at $r=5~\mathrm{kpc}$. The magnetic field lines point along the dividing lines and therefore have direction $\hat b = \sin(i)\hat r + \cos(i)\hat\phi$. 
The magnetic flux is conserved because the magnetic field strength falls off with $1/r$ and the overall flux of all spiral arms are summed up to zero: $\sum_{j=1}^8 f_j b_j = 0$ to ensure zero divergence of the spiral disk field.  
Thereby the numbers $f_j$ reflect the relative cross-sectional areas. The spiral structure of the disk field is defined for $5~\mathrm{kpc} \leq r \leq 20~\mathrm{kpc}$. 
The exact values of the constants $r_{-x}$ and $f_j$ of this formulation can be found in \cite{jf12a}.
The inner part ($3~\mathrm{kpc} \leq r < 5~\mathrm{kpc}$) is modelled as a ``molecular ring'', i.e. a purely azimuthal field component with strength $b_\mathrm{ring}$.

The above description is only valid for the disk component. Therefore, this field component is limited by a height parameter $h_\mathrm{disk}$ and a transition width $w_\mathrm{disk}$, which defines the transition between the disk and halo field according to 
\begin{equation}
 L(z,h,w)=(1+e^{-2(|z|-h)/w})^{-1}.
\end{equation}
The disk field is multiplied by $(1-L(z,h_\mathrm{disk},w_\mathrm{disk}))$ where the halo has a pre-factor of $L(z,h_\mathrm{disk},w_\mathrm{disk})$.

The large-scale disk field formula has  11 parameters: the strength of the toroidal molecular ring region (0.1~$\mu\mathrm{G}$), eight parameters for the field strength of the spiral arms at 5 kpc (0.0 to 4.2~$\mu\mathrm{G}$), and two parameters for the transition scales from disk to halo.

\subsubsection{Toroidal halo field}
Two halo field components are defined by JF12: a purely toroidal halo component and an out-of-plane component, also denoted as the X-shaped field. The field vectors of both components are perpendicular to each other.

The field strength of the purely toroidal field component is characterized by an exponential scale height $z_0$ and magnitudes for the North and the South half of the halo, respectively. The transition between halo and disk is modelled as above:
\begin{equation}
B_\phi^{tor}(r,z)=e^{-|z|/z_0} L(z,h_\mathrm{disk},w_\mathrm{disk}) \cdot B_X \cdot (1-L(r,r_\mathrm{X},w_h)).
\end{equation}
Here $X$ stands for the northern and southern parameters respectively. This formulation gives a field structure for the halo that differs from the structure of the disk and allows for different field strengths and extents in the north and in the south.
The toroidal halo field is modelled by six parameters: two parameters for the amplitudes (1.1 and 1.4~$\mu\mathrm{G}$), and four parameters for the transition in radial and vertical direction.

\subsubsection{X-shaped poloidal halo field}
Observations of external edge-on galaxies \cite{krause09,beck09} motivate the modelling of an out-of-plane halo component. 
The out-of-plane `X' component is characterized by being axisymmetric and purely poloidal. 

The field lines of this component run from the southern direction at a specific angle $\Theta$ towards the $z$-axis; subsequently they cross the $xy$-plane and leave the plane at the same angle. This leads to the name  `X'-shape.
The field strength at any point is equal to the value at the point of the $x$-$y$-plane where the field line crosses the plane. 
The field in the plane is given by a central value at the origin and a characteristic exponential scale length. 
The field strength  is then fully determined by the condition $\nabla\cdot \mathbf B = 0$. The field lines are oriented in the poloidal direction with an angle $\Theta$ between $z$-axis and the field line which depends on radial position. Thus the elevation angle is constant in the outer region and changes in the inner region from this constant angle to zero towards the centre. 

The poloidal halo field is modelled by four parameters: a central field strength (4.6~$\mu\mathrm{G}$), an elevation angle for the outer region, and two radial parameters which give the radius where the elevation angle starts to change, and an exponential scale length.

\subsection{JF12 small-scale models}\label{sec:JF12small-scale}
In addition to the large-scale field components in \cite{jf12a} an anisotropic random field is introduced (which is identified by the term `striated' random field in their work). This anisotropic random field is fully aligned with all components of the large-scale field (i.e. spiral disk and both halo field components).
More precisely, the strength of the anisotropic random field is coupled to the strength of the large-scale model of JF12 by a scalar parameter. 
Furthermore \cite{jf12b} complement their model with a purely isotropic random field.
This  is independent of the large-scale model and its spiral disk strengths and halo components are  determined separately.
The random field disk has a constant field strength in the inner 5 kpc and a spiral structure in the outer part, with the same form as the large-scale spiral disk but a different field amplitude in each spiral arm. The field also scales as $1/r$. 
The random halo field decays exponentially in the radial direction and is Gaussian in the $z$-direction: $B_\mathrm{halo}=B_0 e^{-r/r_0} e^{-z^2/z_0^2}$.
The root-mean-square (RMS) random field strength is then defined as 
\begin{equation}
B_\mathrm{rand,JF12} = \sqrt{B_\mathrm{disk}^2+B_\mathrm{halo}^2}.
\end{equation}
The JF12 isotropic random field model possesses 13 parameters: 8 values for each spiral field strength (ranging from 2 to 37~$\mu\mathrm{G}$), one for the inner part of the disk (7.6~$\mu\mathrm{G}$), and one for the Gaussian scale height of the disk. The halo has three parameters: One central field strength (4.7~$\mu\mathrm{G}$) and two values which control the radial exponential decay and give the Gaussian scale height. 

In the JF12 model for the polarized intensity only the regular and the anisotropic random field have to be taken into account.
Therefore the following simple addition is performed, because the anisotropic random field is aligned with the large-scale regular field with an amplitude of $B^2_\mathrm{striated} = \beta \cdot B^2_\mathrm{reg,JF12}$ and the PI is computed by replacing
\begin{equation}B^2_{\mathrm{reg,\perp}} \rightarrow \alpha(1+\beta) B^2_{\mathrm{reg,JF12}}\sin^2 \theta.\end{equation}
where $\theta$ is the angle between the integration line-of-sight (LOS) and the vector of the large-scale magnetic field with $B^2\sin^2\theta = B_\perp^2$.
We use the original parameters $\gamma=\alpha (1+\beta)=2.92$ and $\beta=1.36$ from the JF12 model. $\beta$ describes the strength of the anisotropic random field component.
$\alpha$ is a scaling factor for the cosmic-ray electron density $n_\mathrm{CRe}$, because in the optimization process the strength of the small-scale magnetic field and the rescaling of $n_\mathrm{CRe}$ are degenerate.
Due to the difference in the new GALPROP model $n_\mathrm{CRe}$ is upscaled by a factor of 1.4. This would correspond to using $\alpha = 1.73$.

The effect  on the total synchrotron intensity produced by the (an)isotropic random magnetic field is accounted for by replacing $B^2_\mathrm{reg}$ with
\begin{equation}B^2_{\mathrm{reg},\perp} \rightarrow \alpha (1+\beta)B^2_{\mathrm{reg,JF12}}\sin^2\theta + \frac 23\alpha B^2_{\mathrm{rand,JF12}}.\label{eqn:JF12random}\end{equation}
In the JF12 model the ensemble of the random magnetic field averages to $\langle B^2_{\mathrm{rand},\perp} \rangle = 2/3 \cdot B^2_{\mathrm{rand}}$, which is accounted for by the second term of Eq. (\ref{eqn:JF12random}) and is independent of the angle $\theta$.
$B_{\mathrm{rand,JF12}}$  is a scalar amplitude giving a large-scale distribution of the strength of the isotropic random field.

The various components of the small-scale magnetic field contribute differently to each of the observables (e.g. the anisotropic random field contributes to I and PI, and the random field only contributes to I).
In the analysis of JF12 an ensemble average is computed analytically. Subsequently, the computed ensemble average is compared to the existing realization without explicitly taking into account any variance.
Therefore there is no influence of the small-scale model on  the RM or UHECR deflection maps in the JF12 model.

\subsection{Modifying the JF12 small-scale field models}\label{sec:modifyingJF12}

The two prescriptions introduced above for replacing $B_{\mathrm{reg}}^2$ in the original formulation of the JF12 model only allow  calculating ensemble averages for the random field.
For example, in the  integral along the LOS for the RM ($\sim \int_{\ell}  n_e (B_\|+b_\|) dl$) the term $b_\|$ for the small-scale magnetic field is not negligible and does not average out in the simulated RM map nor in the map of the  magnetic field, although it plays no role for the ensemble average value. In our approach we use this additional information of $b_\|$, which is implicit in the RMs, to characterize the small-scale magnetic field fluctuations.
Since we want to allow for larger fluctuations, it is not sufficient  only to  compute the ensemble average and so we improve their previous description of the total magnetic field by introducing individually computed realizations of the small-scale field in order to be able to account for variations induced by the computed realizations

\begin{equation}\mathbf{B}_{\mathrm{tot}} = \mathbf{B}_{\mathrm{reg,JF12}} + f_a\mathbf{B}_{\mathrm{aniso}} + f_i\mathbf{B}_{\mathrm{iso}}.\end{equation}
This modification holds for all observables and no additional assumptions on the influence on different observables are necessary besides the usual LOS integrations. 
The parameters $f_i$ and $f_a$ are introduced to allow for a rescaling of the JF12 random field strengths.
The contributions of the (an)isotropic field components are now vector quantities with a functional form as follows.
The isotropic random field is realized by 

\begin{equation}\mathbf{B}_{\mathrm{iso}} = B_{\mathrm{rand,JF12}}\cdot\mathbf{\mathcal{G}}\end{equation}
where $B_{\mathrm{rand,JF12}}$ gives a scalar amplitude of a large-scale distribution of the isotropic random field. It consists of a spiral disk and a halo component.
$\mathbf{\mathcal{G}}$ is a three dimensional divergence-free Gaussian random field with $\langle \mathbf{\mathcal{G}} \rangle=\mathbf{0}$ and $\langle|\mathbf{\mathcal{G}}|^2\rangle=1$ following an arbitrary power distribution of $k$ in Fourier-space.
We incorporate the algorithms of the GARFIELDS code \cite{kitaura08} into HAMMURABI for the vector computation of the Gaussian random fields.
We generate a stochastic realization of such a Gaussian random field with a power spectrum $P(\mathbf{k})$ with the help of an inverse Fourier transformation (see e.g. \cite{martel05}),  written in general as

\begin{equation}B(\mathbf{x}) \propto \int \sqrt{P(\mathbf{k})}(\chi-i\zeta)e^{(i\mathbf{k}\cdot \mathbf{x})}d\mathbf{k},\end{equation}
where $\chi$ and $\zeta$ are random numbers drawn from a Gaussian random distribution. The $\mathbf{k}$-vectors are then transformed to real-space by a discrete fast Fourier transformation\footnote{http://www.fftw.org.}.
Here $k=2\pi/l$ is the wave vector and $l$ the spatial scale.
The anisotropic random field is modelled by

\begin{equation}\mathbf{B}_{\mathrm{aniso}} = \sqrt{\frac 32\beta}\cdot \frac{\mathbf{\mathcal{G}}\cdot \mathbf{B}_{\mathrm{reg,JF12}}}{|B_{\mathrm{reg,JF12}}|} \mathbf{B}_{\mathrm{reg,JF12}}.\end{equation}
This formulation ensures that the resulting magnetic field vector has an orientation which is aligned along the large-scale field vector of the JF12 model $\mathbf{B}_{\mathrm{reg,JF12}}$ and is parallel or anti-parallel to the large-scale field.
The pre-factor $\sqrt{3/2}$ compensates the reduction of $\mathbf{\mathcal{G}}$ by one dimension due to the projection.
Both the sign and strength of the field vary on small scales.
This formulation produces a anisotropic random field as the description in JF12 suggests. Therefore the anisotropic random field follows the large-scale field structure and has a spiral disk and poloidal and toroidal halo components.

We assume the power distribution of the Gaussian random field $\mathbf{\mathcal{G}}$ to be of Kolmogorov type.
The lower limit for $k$ is given by $k_\mathrm{min} = 2\pi /L_\mathrm{max}$ and the upper limit of $k$ is determined by the spatial resolution at the Nyquist limit $k_\mathrm{max} = 2\pi / L\cdot N/2$, where $L$ is the box size and $N$ the number of grid cells in such a box.

The typical length scale of our small-scale field is of the order of  hundreds of parsecs.
The correlation length $L_c$ of a Gaussian random field can be defined (see Eq. (2.4) in \cite{harari02}) as
\begin{equation}L_cB^2_{\mathrm{rms}}=\int_{-\infty}^{\infty} \langle\mathbf{B}(0)\cdot\mathbf{B}(\mathbf{x}(l))\rangle ~\mathrm{d}l,\end{equation}
where $\mathbf{x}(l)$ is a point in space displaced by distance $l$.
Eq. (2.6) of \cite{harari02} gives the following formula for computing the correlation length $L_c$ for a Gaussian random field which is sampled between $k_{min}=2\pi/L_{max}$ and $k_{max}=2\pi/L_{min}$ with a power law with slope $n$
\begin{equation}L_c = \frac 12 L_{max} \frac{n-1}{n}\frac{1-(L_{min}/L_{max})^{n}}{1-(L_{min}/L_{max})^{n-1}}.\end{equation}
A Kolmogorov spectrum ($n=5/3)$ results in $L_c \approx L_{max}/5$ for broad spectra.

We compute both random components from a three dimensional realization of a spectral power distribution. The spatial resolution throughout the whole sampled volume is 20~pc.  The minimum length scale $L_{min}$ therefore is 40~pc due to the Nyquist limit.


\section{Results}

We use the HEALPix pixelization scheme by \cite{gorski05}, with a resolution parameter of $\mathrm{NSIDE}=128$,
so that the angular resolution of the final maps is approximately 27 arcmin.
First each HEALPix-map pixel is sampled along the LOS into a `3D-HEALPix grid' as suggested by \cite{waelkens09}.
This results in a conical shape of the volume cells along this line, i.e. the volume units increase with distance $l$ along the LOS.
The resolution is enhanced by splitting one such beam into 4 sub-beams and is performed repeatedly in such a way that the spatial resolution is kept similar to the resolution of the random field grid.
The value for one observational pixel then arises from the contribution of multiple sub-beams, which approximately allows us to take into account effects such as beam depolarization (see Appendix A.1 of \cite{waelkens09} for details).
This is required because sub-beams undergo different amounts of Faraday rotation.
 
We simultaneously compare all computed observables to the observational data.
We follow \cite{oppermann12} in our analysis of the RM data, which we take from the updated full-sky map of \cite{oppermann14} and we denote the observational data as O12.

\subsection{Synchrotron and RM full-sky maps}

\begin{figure}
\begin{center}
  \includegraphics[width=0.99\textwidth]{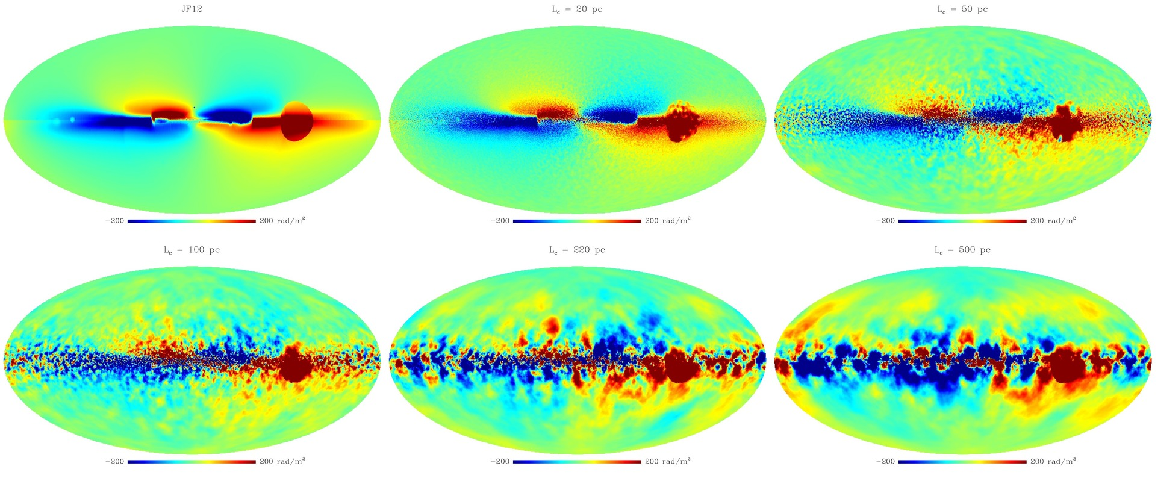}
  \caption{Comparison of simulated RM maps for various length scales of the turbulent magnetic field.
  It can  be seen that the larger the length scale of the turbulent field, the more prominent it appears.}
  \label{fig:rm_lc}
\end{center}
\end{figure}

In order to give a basic measure for the quality of the model, a mean (weighted) quadratic deviation of simulated data points ($sim$) from observational data points ($obs$) is estimated as follows,
\begin{equation}
\Delta^2 = \frac {1}{n_{w}} \sum_i w_i(\mathrm{data}_i^{\mathrm{sim}} -\mathrm{data}_i^{\mathrm{obs}})^2,
\end{equation}
where $w_i$ is a weight for the data point $i$ and $n_{w}=\sum_i w_i$ is the norm.

\begin{figure}
\begin{center}
  \includegraphics[width=0.475\textwidth]{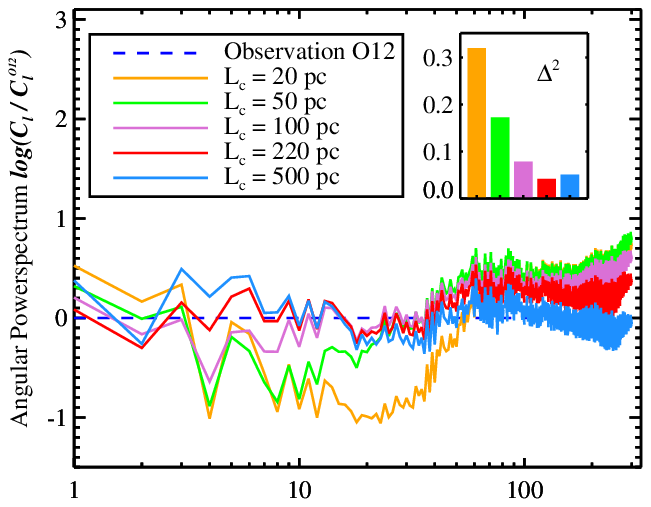}
  \caption{Comparison of an ensemble of simulated RM signal map power spectra (solid lines) for various length scales of the turbulent magnetic field power spectrum.
  We normalise all the mean power spectra to values derived from the O12 data (dashed line and Fig. \ref{fig:rescaled}).
  We find the best fit at a correlation length of 220~pc which corresponds to a maximum length scale of 1 kpc.
  We also show the quadratic deviation $\Delta^2$ of the angular power spectra from the observational curve.
}
  \label{fig:spec}
\end{center}
\end{figure}

First we adapt the amplitudes of the two small-scale magnetic field components of the JF12 model.
The new prescription for the small-scale magnetic field leads to contributions of the small-scale fluctuations to the full sky maps, which are not present in the JF12 model.
They appear because larger scales are allowed and these do not average out, because too few large-scale fluctuations along the LOS are present (see Fig. \ref{fig:rm_lc}).
This is illustrated in Fig. \ref{fig:spec}, which shows that the larger the maximum length scale of the random field $L_{\textrm{max}}$, the more power is present in the full sky map at larger angular scales (smaller $l$).
We calculate a mean quadratic deviation of the data points in the angular power spectrum plot: 
$\Delta^2 = \sum_l w_l (\log C_l - \log C_l^\mathrm{O12})^2 / n_{w}$. Thus the data are weighted by 
the inverse of the data point density on the x-axis: $w_l = \log(1+1/l)$.
We find the minimum difference of observed and simulated power spectrum curve (and therefore minimum $\Delta^2$) for a value of $L_{c}$ = 220~pc (red data), which therefore best fits the observational data.

\begin{figure}
\begin{center}
  \includegraphics[width=0.475\textwidth]{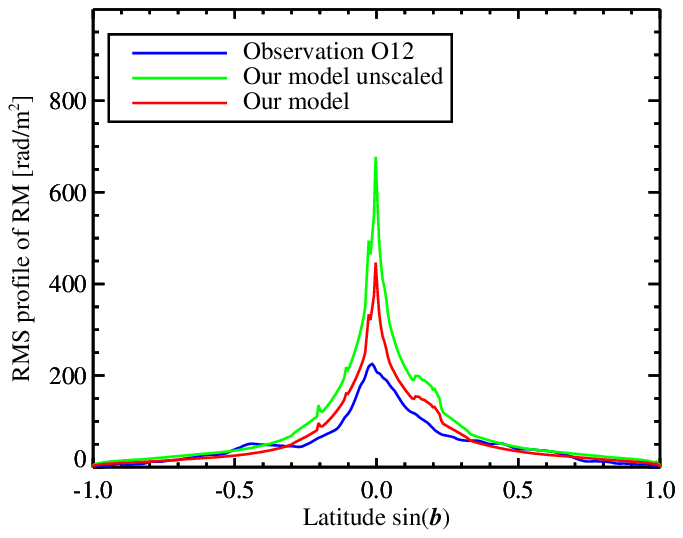}
  \caption{RMS profiles of RM distributions as a function of Galactic latitude.
  Profile derived from the RM data of \cite{oppermann12} (blue line), and  profiles corresponding to our new prescription with different scaling values for $f_i$.
  The green line corresponds to the JF12 parameter values ($f_i=f_a=1$) and the red line corresponds to our best fit values.}
  \label{fig:profile}
\end{center}
\end{figure}

We follow \cite{oppermann12} and compute the root mean square (RMS) profiles of the RM maps as a function of Galactic latitude $b$.
As we show in Fig. \ref{fig:profile} an unscaled small-scale magnetic field leads to a dominant contribution of the random magnetic field and large-scale features of the RM map vanish (not shown).
We set $f_i$ to 0.6 and $f_a$ to 0.3 to rescale the amplitude of the random magnetic field and obtain the best-fit to the observed maps, and a RMS profile curve which also fits  better to the data. Although the latitude profile does not perfectly fit to observations, we at least improve it by rescaling but clearly find best fitting for all other observables.
\begin{figure}
\begin{center}
  \includegraphics[width=0.475\textwidth]{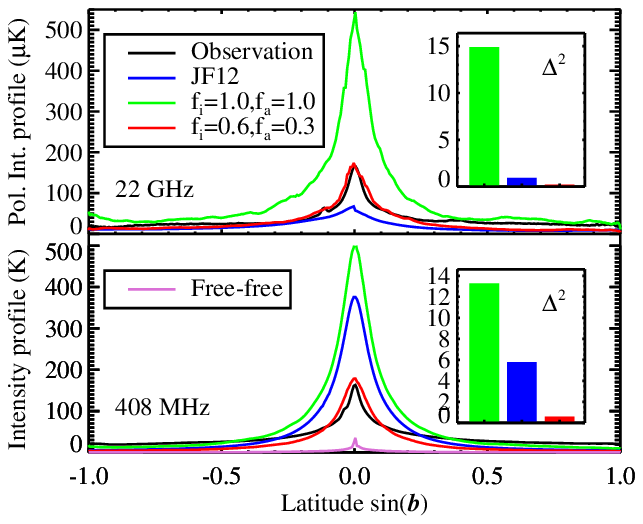}
  \caption{Latitude profile of the mean synchrotron power. The panels show the profiles for total synchrotron intensity at 408 MHz and polarized intensity at 22 GHz. 
  The profiles of the simulated maps are compared to the observational data. The inset shows the mean quadratic deviation of the curves from the observational curve. See text for details.}
  \label{fig:profiles_synchrotron}
\end{center}
\end{figure}
In Fig. \ref{fig:profiles_synchrotron} latitude profiles for total synchrotron and polarized synchrotron emission at 408 MHz and 22 GHz are shown for the observational data, the original formulation of the JF12 model and for our (un)scaled formulation. The profile curves show that the choice of $f_i$ and $f_a$ significantly improves the agreement for the latitude profiles, where we estimate the mean quadratic deviation $\Delta^2$ similarly to the  description above.
We find a significant improvement of $\Delta^2$: the value decreases (which means a better fit) from 5.80 to 0.60 for the total intensity profile at 408 MHz. The $\Delta^2$   for 22 GHz polarized emission decreases from 0.95 to 0.22.

The JF12 parameters (including the large-scale ones) have to be re-adjusted due to the fact that thermal free-free emission is not accounted for in the classic JF12 model, since we put special emphasis on the small-scale component.
We estimate the contribution from free-free to be about 10-20\% of the total emission (Fig.~4 in \cite{orlando13}). In Fig. \ref{fig:profiles_synchrotron} we additionally show the free-free emission profile at 408 MHz, which is based on the prescription of \cite{rohlfs96}.  The spatially varying thermal electron temperature $T_e$  used  is given by \cite{sun08} based on the work of \cite{quireza06}, \cite{reynolds99}, and \cite{peterson02}: $T_{e,\mathrm{model}}(\mathbf{r})= 5780 +287r-526|z|+1770z^2$, where
$T_e$ is in units of Kelvin and the distances are in kpc. This electron temperature distribution model was optimized for and compares well to the WMAP free-free template for 22 GHz and was used to calculate the emission at 408 MHz. HAMMURABI also accounts for free-free absorption.

\begin{figure}
\begin{center}
  \includegraphics[width=0.475\textwidth]{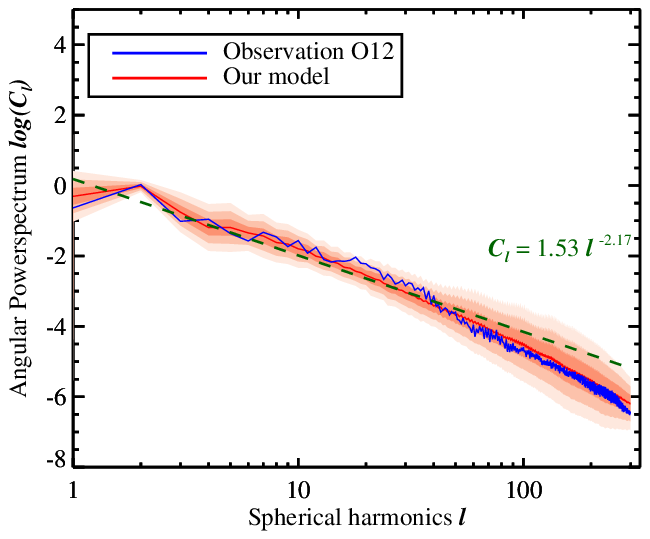}\vspace{0.2cm}
  \caption{Power spectrum of the signal map corresponding to the RM map of \cite{oppermann14} (blue line).
  Also plotted is the mean power spectrum of the signal map derived from our new model (red line), where the contour levels represent  standard deviations of $1\sigma$, $2\sigma$ and $3\sigma$.
  We derived the mean and standard deviations from a sample of 100 simulated realizations of the turbulent magnetic field power spectrum.}
  \label{fig:rescaled}
\end{center}
\end{figure}

Fig. \ref{fig:rescaled} shows that after rescaling the amplitude of the random field the shape and quality of the angular power spectrum is  preserved.
Since the observed map results from one realization of infinitely many possible ones, we compare the spectrum of the observed map to a sample of 100 realizations.
We show the mean values as well as confidence levels for $1 - 3$ standard deviations $\sigma$.

\begin{figure*}
\begin{center}
  \includegraphics[width=0.95\textwidth]{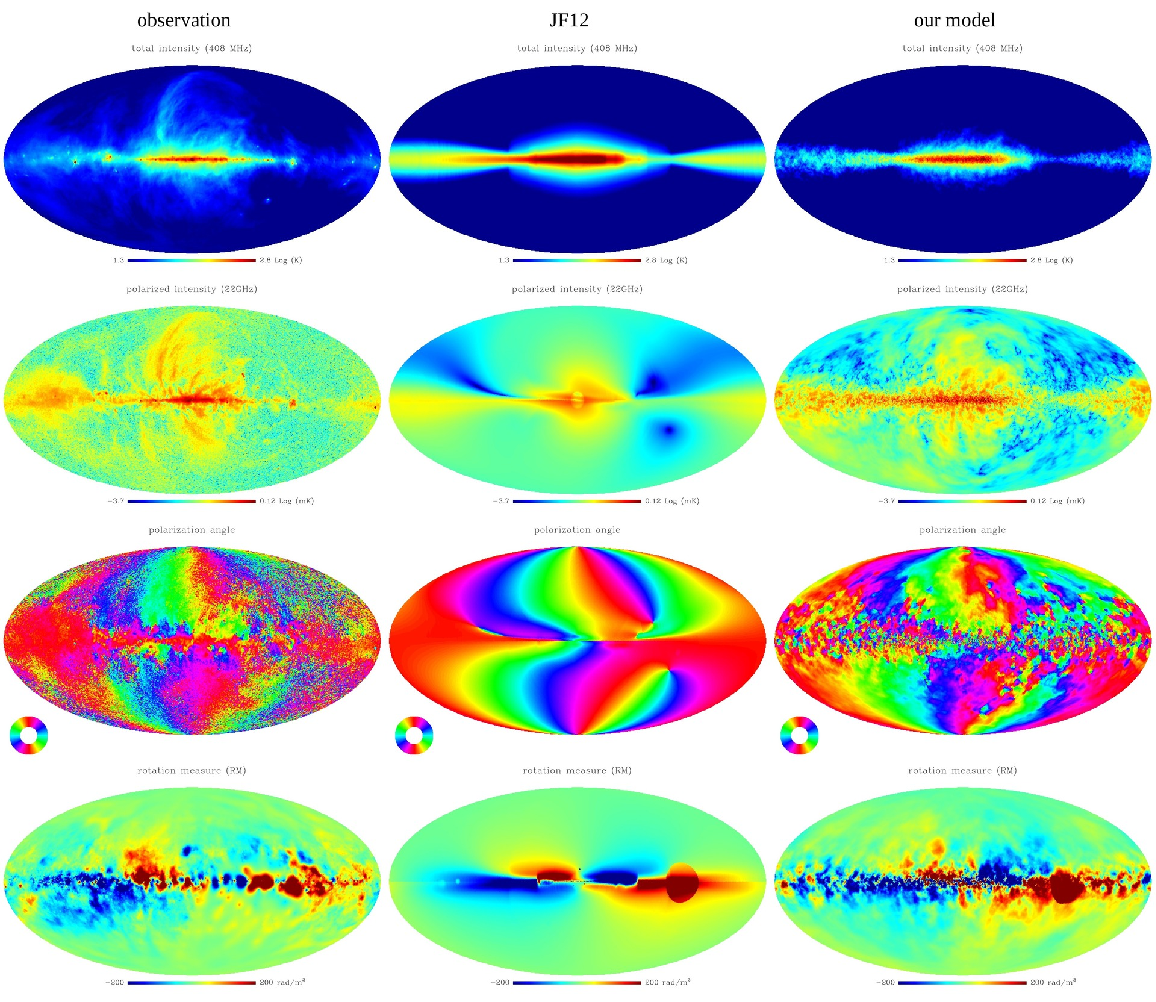}\vspace{0.2cm}
  \caption{Mollweide projections of observations and simulated data.
  In each of the maps, the Galactic longitude is zero in the centre and increases to the left.
  From top to bottom: the total synchrotron intensity, the polarized synchrotron intensity, the polarization angle and the RM distribution.
  From left to right: the observed quantities, the JF12 model predictions and our new model predictions.
  For the comparison we use the total 408 MHz intensity data of \cite{remazailles14}, the polarized intensity and polarization angle as given by WMAP 22 GHz observations and the RM map of \cite{oppermann14}.
}
  \label{fig:compare}
\end{center}
\end{figure*}

Fig. \ref{fig:compare} shows a full-sky comparison of predictions of our model, the JF12 model and observational data.
We compare the total synchrotron intensity, the polarized synchrotron intensity, the polarization angle as well as the rotation measure distribution.
Our model compares very well with the observational data for the large-scale features of the map and is also in good agreement  for the characteristics of the small-scale features.
In contrast the JF12 model does not produce small-scale structures in the full-sky maps, because it was only constructed to reproduce the large-scale features. With our formulation we are capable to explicitly treat the small-scale fluctuations.

We now discuss the contribution of the different magnetic field components to the observables. 
The Haslam 408 MHz data \cite{haslam81,haslam82,remazailles14} gives a good measure  of the foreground synchrotron emission.
In this total intensity data the perpendicular projections of all components of the magnetic field contribute to the synchrotron emissivity: large-scale field components and both small-scale components, i.e. anisotropic and isotropic random field. 
We find good agreement with the overall large-scale structure of the map for our model.
The large-scale structure, which was fitted in JF12, is not smoothed out by an individual realization of the random component.
It is clear that the map of the JF12 model with the synchrotron contribution as described in Sec. \ref{sec:modifications} does not produce any small-scale features in the synchrotron maps (due to the smooth distribution of the cosmic ray electrons).
We note that the overall amplitude of the total intensity in the map of the JF12 model gives slightly too much emission compared to observations.
We also compare to the polarized synchrotron map of the WMAP 7 22~GHz data.
In the JF12 model only the large-scale and anisotropic random magnetic field components contribute to the polarized intensity, but in our model this map is sensitive to all the magnetic field components as already mentioned.
In our model we find good agreement with the large-scale structure as well as small-scale features which are not reproduced by the JF12 model.
The features in the PI 22GHz map which are in the region of Loop I result from the large-scale JF12 model in combination with our formulation of the anisotropic random field, which gives larger structures which are parallel to the large scale magnetic field model. 
The JF12 model gives high emission in Stokes Q and U in the region which is identified by Loop I. So direction and strength are incorporated into the modelling process for this feature. 
Due to the combination of the JF12 B-field distribution and our formulation of the small-scale field (especially the larger coherence length), structures in the polarized synchrotron emission become evident. 
If it was desired to describe the emission of Loop I more accurately, it should be modelled explicitly.
We find a similarly good agreement of the polarization angle for the halo field. The polarization angle of the disk emission in the anti-centre direction is not well reproduced. The fluctuations of our models give deviations from zero polarization angle. Hence they seem to be slightly overestimated in the outer disk.

Finally we compare with the RM map of \cite{oppermann14}.
In the JF12 model the RM is unaffected by small-scale magnetic field structures because they compute the ensemble average for each pixel, which means that their analysis neglects the fluctuations in the RM values.
We note that the map of the JF12 model  only gives some localised small-scale features in the RM map which result from the implementation of local features in the NE2001 model for the thermal electrons, but not from the magnetic field model itself.
Again we find good agreement of our model with the large-scale structure as well as with the small-scale structures, which is substantiated by the spectral analysis of this map (see above). 

\begin{figure}
\begin{center}
  \includegraphics[width=0.66\textwidth]{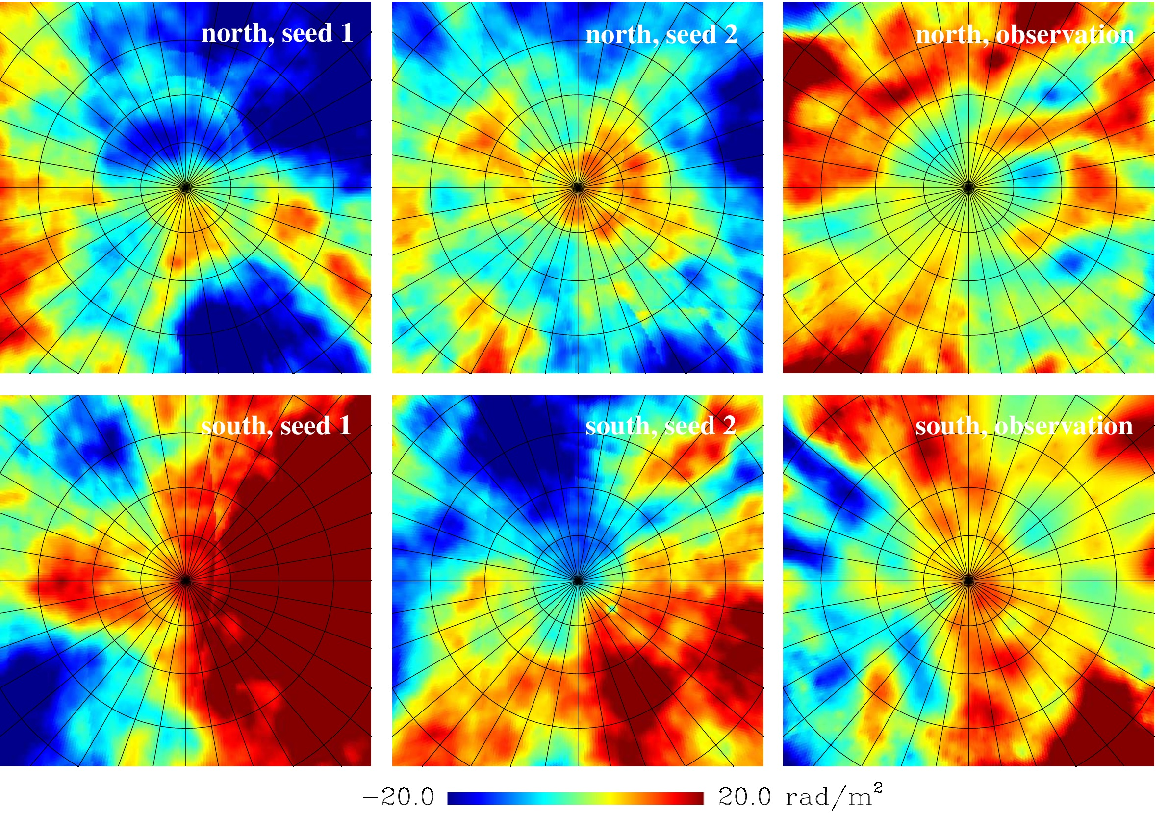}
  \includegraphics[angle=90, width=0.33\textwidth]{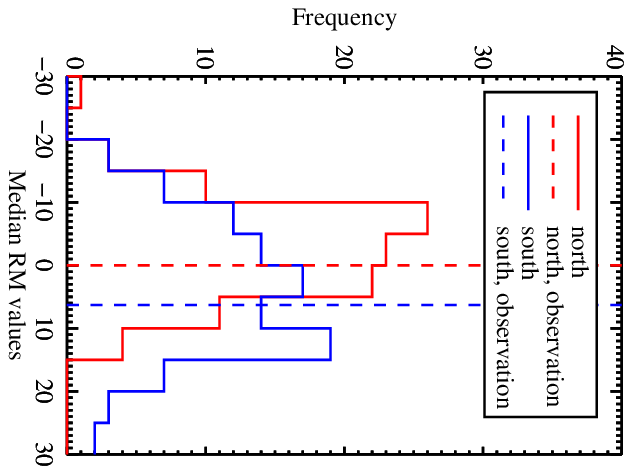}
  \caption{Computed RM maps towards the north and the south directions. The grid spacing is 10$^\circ$. The two left columns show the maps for different field realizations. This illustrates that the RM values towards the poles are strongly dependent on the choice of realization.
For 100 of these realizations we find median values of $-2.8\pm7.1$ rad/m$^2$ and $4.4\pm11.0$ rad/m$^2$ for the northern and the southern direction, respectively. 
The right column of the RM plots shows the observational data set of \cite{oppermann14}. We also show a histogram of the computed median RM values towards the Galactic poles for 100 different realizations. 
The dotted lines show the observational median RM values of more than 1000 polarized radio sources giving values of 0.0 rad/m$^2$ for the north respectively 6.3 rad/m$^2$ for the south Galactic pole \cite{mao10}.}
  \label{fig:rm_north_south}
\end{center}
\end{figure}

In order to investigate the vertical magnetic field towards the Galactic poles the RM data at Galactic latitudes $|b| \ge 77^\circ$ are compared with \cite{mao10} and the RM values at the Galactic poles of \cite{oppermann14}.
In \cite{mao10} more than 1000 polarized extragalactic radio sources have median RM values of $0.0 \pm 0.5 ~\textrm{rad/m}^{2}$ for the north Galactic pole and $6.3 \pm 0.7 ~\textrm{rad/m}^{2}$ towards the south Galactic pole. 
We also calculated the median values for \cite{oppermann14} and found values of 1.9 and 6.7 $\textrm{rad/m}^{2}$ for 2520 HEALPix pixels in these directions respectively and a width of the distribution of 2.8 and 2.9 $\textrm{rad/m}^{2}$. The differences of the results of \cite{mao10} and \cite{oppermann14} are due to the difference in the data sets, i.e. only discrete RM values of extragalactic radio sources or continuous RM maps.
We find median values of $-2.8 ~\textrm{rad/m}^{2}$ and $4.4 ~\textrm{rad/m}^{2}$.
With our model it can be inferred that the RM values for the north and south Galactic poles are dominated by the random field and its realization and is not necessarily restricted to the large scale magnetic field as indicated in Fig. \ref{fig:rm_north_south}. 
The figure also shows the distribution of median RM values in the direction of the poles dependent on the chosen realization. 
The values are broadly spread among the ensemble of realizations of the small-scale field, i.e. $\pm 7.1 ~\textrm{rad/m}^{2}$ and $\pm 11.0 ~\textrm{rad/m}^{2}$ and therefore strongly dependent on the chosen realization. The median values of \cite{mao10} (depicted by the dashed lines) are  0.6 $\sigma$ respectively -0.1 $\sigma$ away from the mean values.
We find that the small-scale field realization that is apparent in our Galaxy could be typical of the ensemble of small-scale realizations which we generated  by our procedure.

We note that the size of the structures in the maps varies over the maps.
For example  for the RM map, we see larger structures for high values of absolute latitude and smaller structures in the disk.
It may be that this is due to the contribution of the thermal electron density $n_e$ along the LOS.
In the disk the angular extent is smaller than in the halo, because of the contribution of structures which are further away (see also similar findings by \cite{sun09}).
In the Galactic plane more small-scale fluctuations add up due to the longer LOS and can even change the sign of the RM.
This is present in the observational RM map as well as in the synthetic map of our model. 

\subsection{UHECR deflection}
\begin{figure}
\begin{center}
  \includegraphics[angle=90, width=0.32\textwidth]{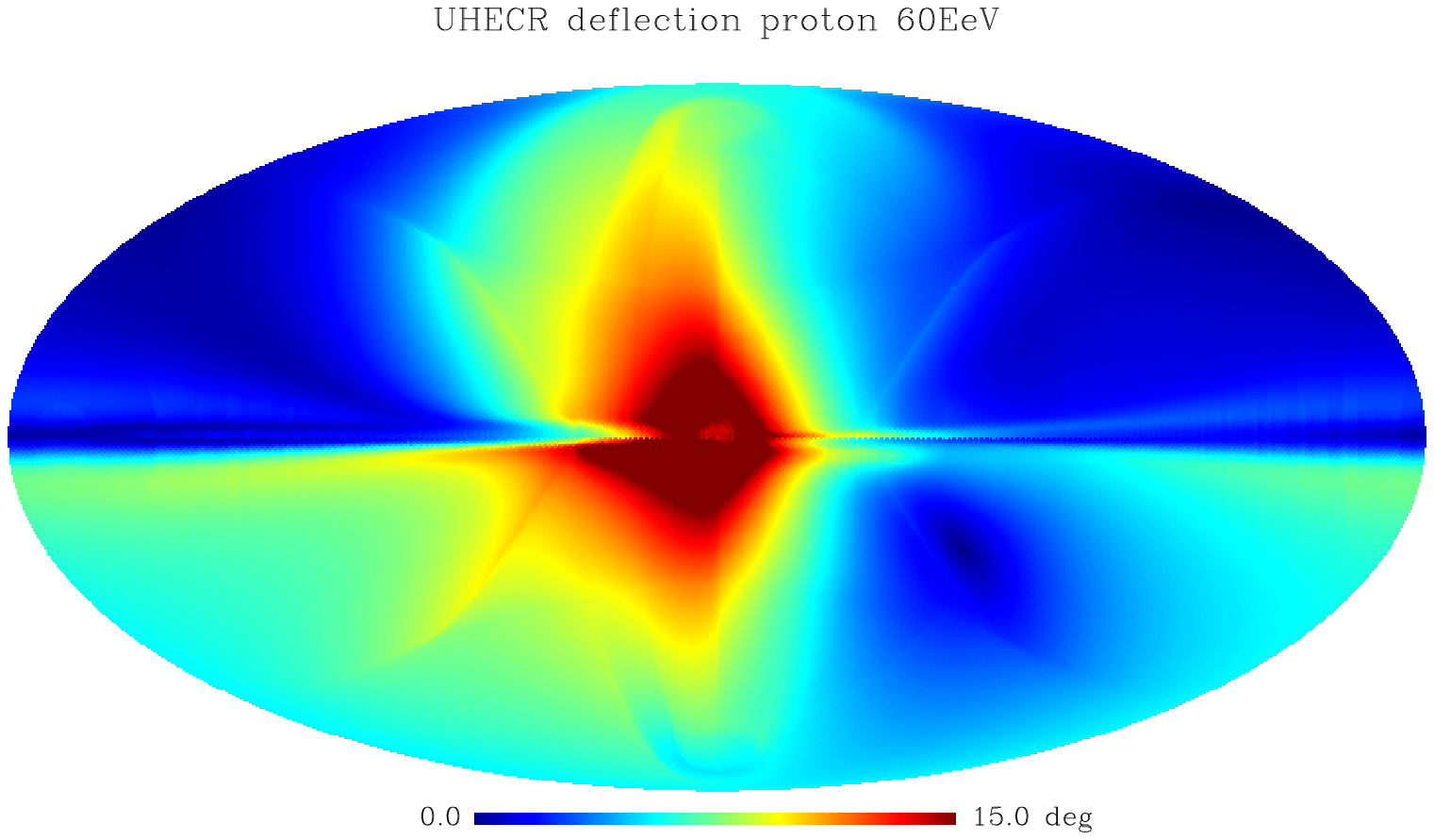}
  \includegraphics[angle=90, width=0.32\textwidth]{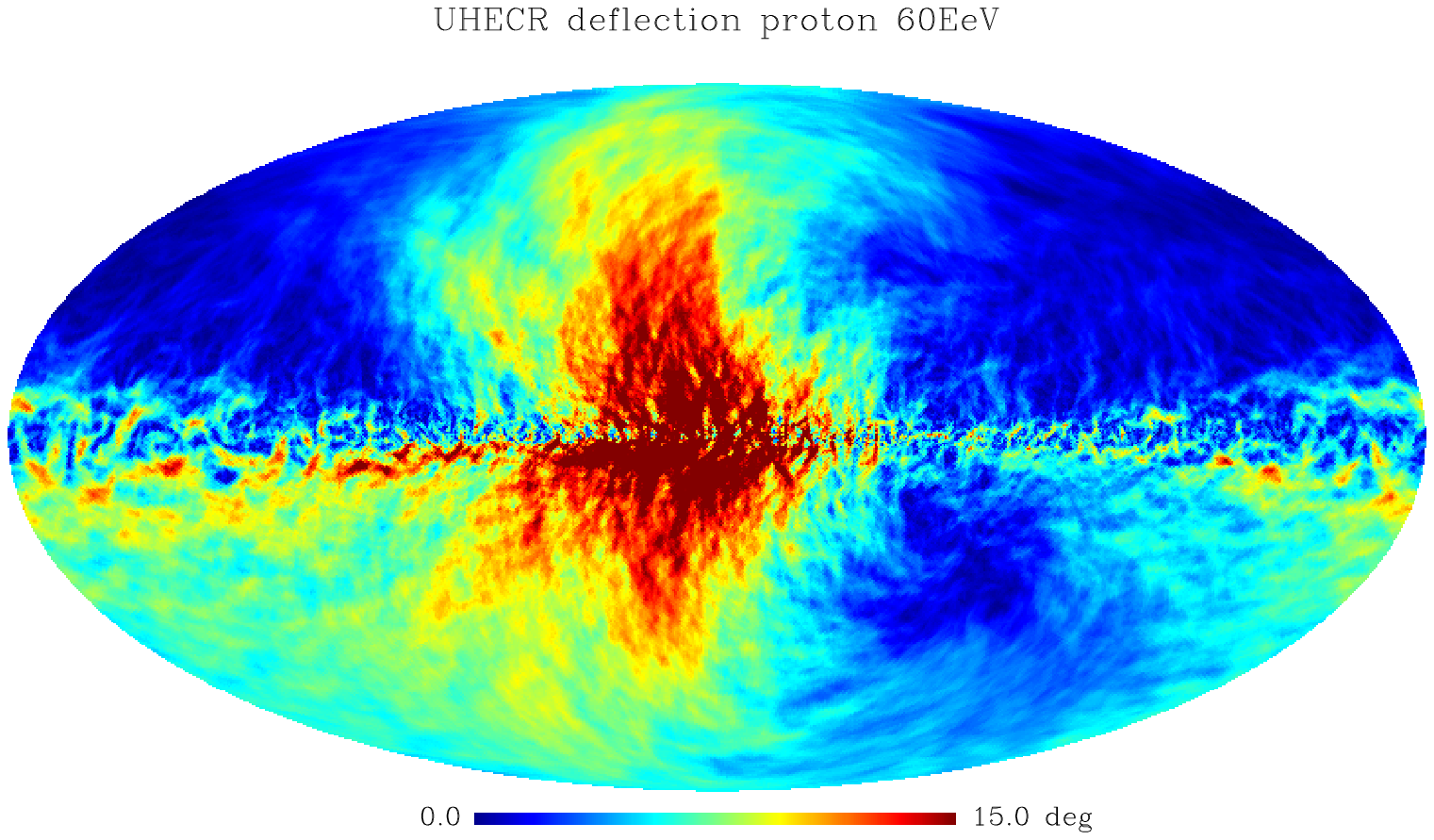}
  \includegraphics[angle=90, width=0.32\textwidth]{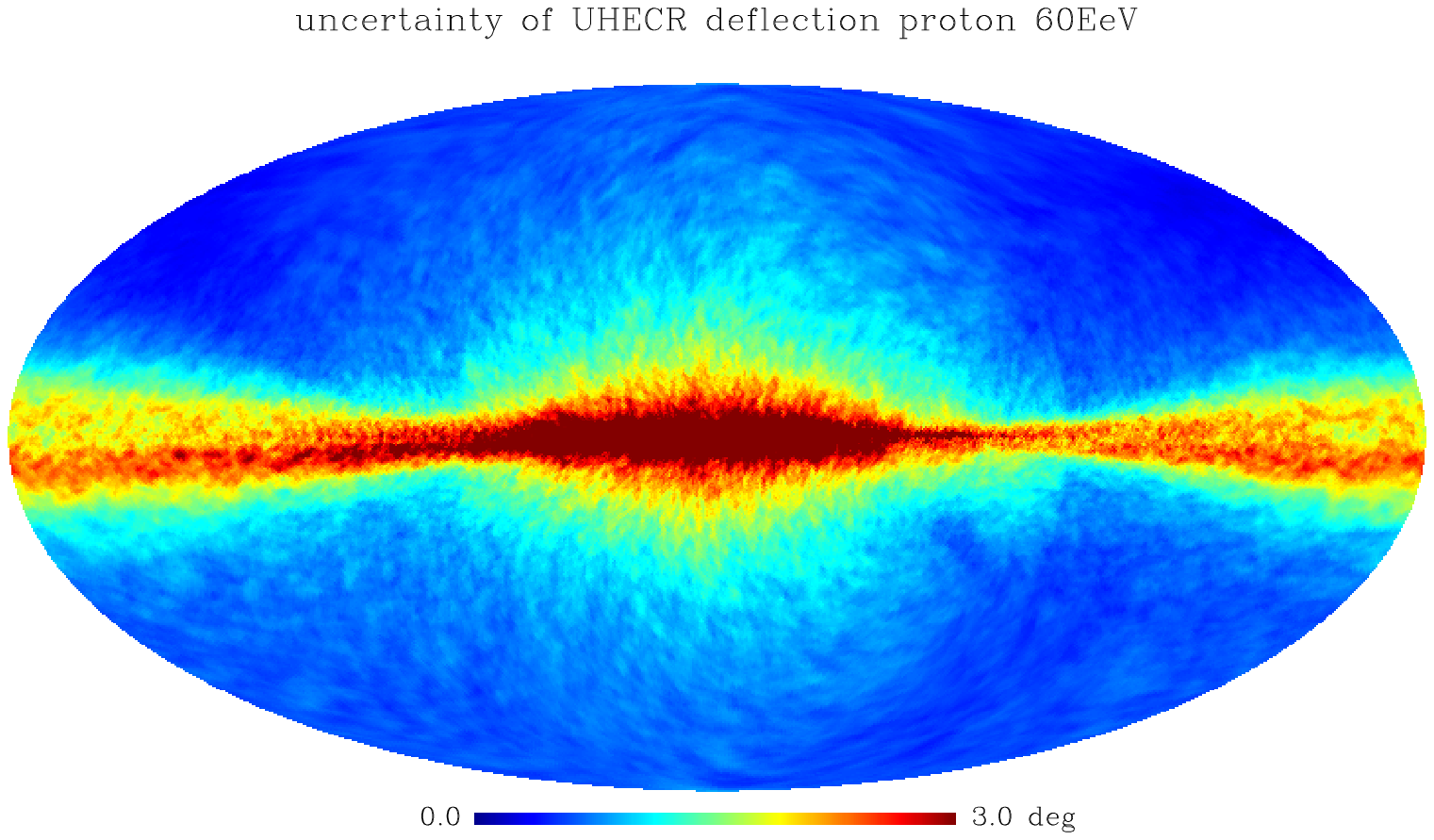}
  \caption{Galactic deflection maps for a 60~EeV proton.
  Absolute magnitude of the deflection, displayed by arrival direction.
  The left panel shows the deflection resulting from the original JF12 model. The middle panel shows the results of one field realization of our model.
  In our new prescription for the GMF we are able to resolve the small-scale features in the deflection characteristics of UHECR.
  The right panel shows a full-sky map of the standard deviation of Galactic deflection maps.
  The per-pixel standard deviation was calculated for a number of 100 different small-scale random field realizations.
  The deflections in the direction towards the Galactic centre and the disk are more affected by the choice of the realization than in the directions towards the Galactic poles.
}
  \label{fig:def_new}
  \label{fig:def_stddv}
\end{center}
\end{figure}

Detailed knowledge of the structure and properties of large-scale and small-scale random fields is needed for many astrophysical purposes.
Extragalactic charged particles are deflected by the GMF. 
For the identification of UHECR sources and the prediction of deflection patterns detailed knowledge of the GMF is essential.
Deflection maps are used which are obtained by approximating the net deflection angle $\Theta$ by computing the integral along the LOS $\ell$ (see e.g. \cite{kachelriess07})
\begin{equation}
\Theta \approx \int_\ell \frac{1}{r_g} dl = \frac{Z q_e}{pc} \int_\ell B_\perp dl
\end{equation}
where $r_g = pc/Z q_e B_\perp$ is the Larmor radius, $Zq_e$ being the charge of particle, $c$ the speed of light and $p$ the momentum perpendicular to the magnetic field $\mathbf{B}$.
Fig. \ref{fig:def_new} shows the deflection map of UHECR protons with energy of 60~EeV for the original JF12 model and for our model.
The overall shape of the deflection map is well preserved, but we find small-scale structures.

\begin{figure}
\begin{center}
  \includegraphics[width=0.475\textwidth]{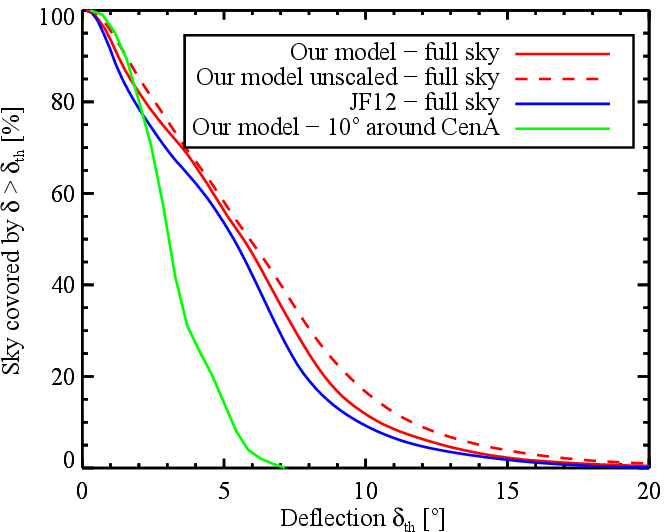}
  \caption{Cumulative fraction of the sky with deflection angle larger than a threshold deflection $\delta_\mathrm{th}$.
  Deflections for our model, the original JF12 GMF model and the deflection of an area of 10$^\circ$ around the direction on the sky pointing to Cen A.
  The  curves are for UHECR with energy 60 EeV.}
  \label{fig:cumul_defl}
\end{center}
\end{figure}

Fig. \ref{fig:cumul_defl} shows the fraction of the sky over which UHECR deflections (for 60~EeV protons) are larger than a threshold deflection angle of $\delta_\mathrm{th}$.
We find that deflections larger than 10$^\circ$ only cover approximately ten per cent of the sky.
It can be seen that our model yields slightly higher values of deflection angles for the single realization that is shown than the JF12 model. The large-scale distribution is conserved, but we find large changes in the local distribution of the deflection angles. 
We quantify the variance in the observables due to the random component in a way that JF12 cannot, because of their insufficient treatment of the small-scale magnetic field.
We conclude that with our model the uncertainties in the deflection angles can be estimated and note that one cannot get around this astrophysical variance unless it would  be possible to map the small-scale component of the magnetic field of the Milky Way in detail.

We find that the small-scale structure of the UHECR deflection maps vary for different realizations.
In order to estimate the influence of choice of realization of the small-scale magnetic field on the UHECR deflection maps, we calculate the standard deviation of each pixel of the deflection map by varying the small-scale random field realizations.
The lower panel of Fig. \ref{fig:def_stddv} shows that the deflection in the central part of the Galactic disk is more strongly affected by the choice of the random field realization. This is due to the larger amplitude of the small-scale field. 
In order to identify sources of UHECRs the investigation of absolute deflection and its uncertainty are important. 
Studying UHECR sources in the plane is difficult, primarily because of the absolute high deflection angles which are not well constrained by GMF models.
But even if an accurate model of the large scale field were available,  one would still have to worry about the problem of the small scale field. Clearly we are far from that situation.

\begin{figure}
\begin{center}
  \includegraphics[width=0.475\textwidth]{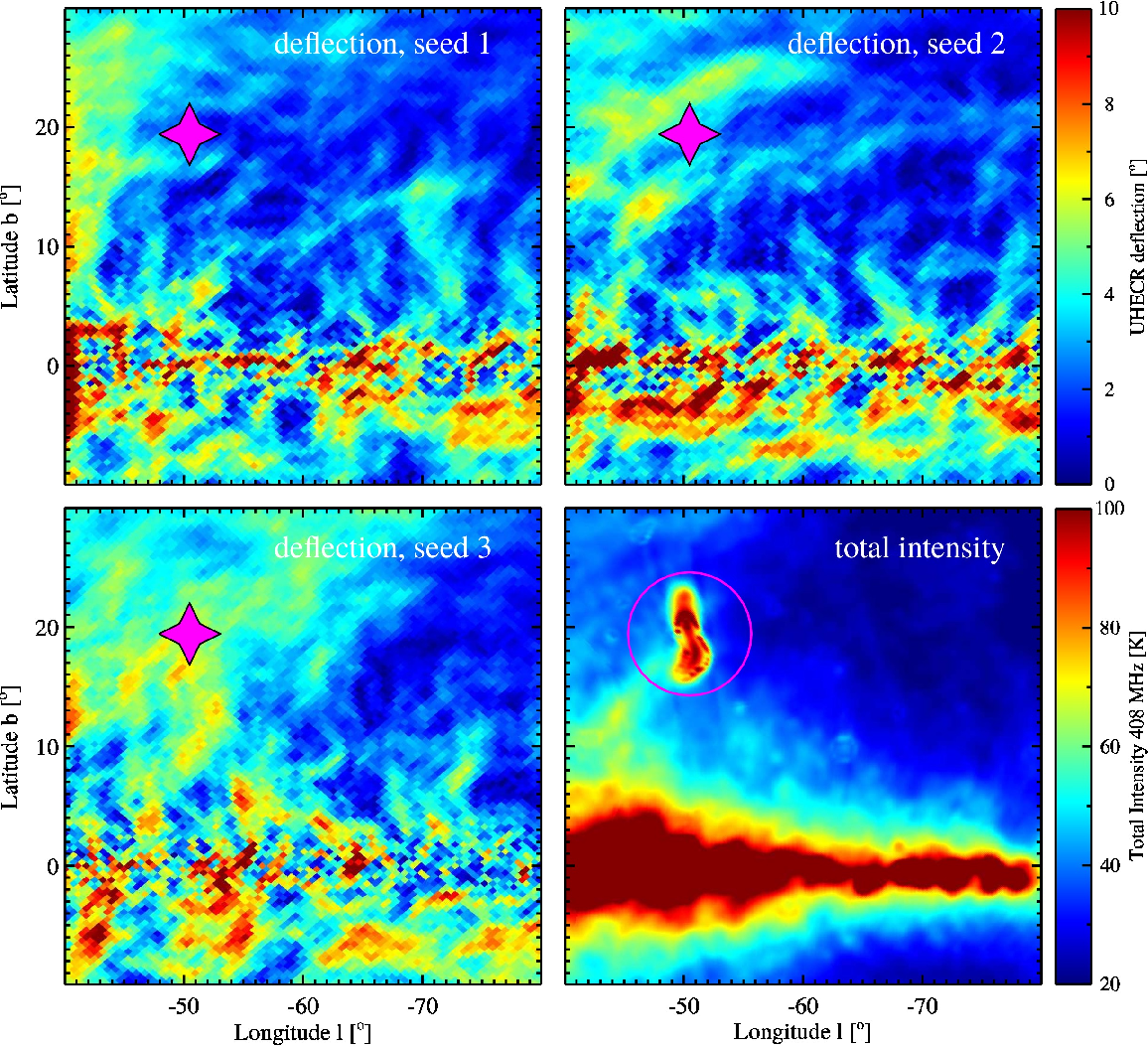}
  \caption{In both upper panels and in the lower left panel we show a zoom-in on the Galactic deflection map (see Fig. \ref{fig:def_new}) 
  for a 60 EeV proton around Cen A (indicated by the purple symbols) for three different random magnetic field realizations.
  We show the absolute magnitude of the deflection, displayed by arrival direction.
  The UHECRs originating from Cen A are deflected by the large-scale field of the JF12 model towards smaller longitudes and lower latitudes.
  The lower right plot shows the same region of the 408 MHz synchrotron intensity map around Cen A \cite{haslam82}.
  The wide extent of the radio lobes can clearly be seen in this version of the total synchrotron map.
}
  \label{fig:def_zoom}  
  \label{fig:haslam_cena}
\end{center}
\end{figure}

In \cite{farrar13} and \cite{farrar14} the deflection of UHECRs in the JF12 model is discussed and compared to other GMF models.
They propose Centaurus A (Cen A) as the most plausible candidate for the origin of UHECRs;
Cen A is the closest active galactic nucleus and is known as a bright radio source (see Fig. \ref{fig:haslam_cena}).
They show the arrival directions of protons assuming different GMF models.
The uncertainties of the arrival direction presented in their article originates from uncertainties in the optimized model parameters.
The deflection which results from any  random magnetic field is not included and only large-scale features are considered. 

In accordance with  \cite{giacinti11} we claim that small-scale features must be taken into account for studies of UHECR deflection.
\cite{keivani15} present simulations of trajectories of UHECRs from Cen A to earth.
They use the JF12 model as well as only the isotropic random component of the JF12 model $B_{\rm rand, JF12}(\mathbf{r})$, but they neglect the influence of the anisotropic random field.
They study the impact of different random field realizations and coherence lengths up to $20 - 100$ pc ($L_{\mathrm{max}}$ up to 512~kpc), but do not discuss their choice of coherence length.
In the present work we are able to bridge the gap and motivate the choice of coherence length.

Fig. \ref{fig:def_zoom} shows a zoom-in on the region around Cen A.
In \cite{keivani15} the source directions of the UHECRs are supposed to lie in the regions of the radio lobes of Cen A (for the wide extent of the lobes see Fig. \ref{fig:haslam_cena}).
Therefore a region of 3 degrees around the center of Cen A is here chosen as the origin of the calculated UHECR trajectories.
The UHECRs are deflected by the large-scale field of the JF12 model towards smaller longitude and lower latitudes.

Fig. \ref{fig:def_zoom} shows the deflection around CenA; as already illustrated in Fig. \ref{fig:cumul_defl} the deflection varies with the choice of the realization of the random fields.
We find a mean value of $\mu\approx3.5\degree$ and a standard deviation of $\sigma\approx\pm1.12\degree$ for the deflection for  a region of the sky of $10^\circ$ around Cen A.

\section{Discussion \& Implications}

Splitting the description of the GMF into a sum of several large-scale and
small-scale components allows to separately investigate the origin of such 
magnetic field components. Especially the turbulent field is of great
interest and its physical origin so far remains unknown. Candidates for the
origin of such small-scale features in the magnetic field are interstellar
turbulence, supernova explosions and supernova remnants, which induce
small-scale distortions in the large-scale field. Such small-scale fields
can interact with shocks and are altered by the dynamics of the gas,
magnetic reconnection or a wide range of complex physical processes.

Regions of supernova and superbubble explosions
are believed to be the main source of energy on similar scales of
about 100~pc. Large loops of radio emission such as the North Polar
Spur or Loop I to IV show the influence of localised features on
magnetic fields and therefore also on the observed maps. Supernovae
which make bubbles in the ionized interstellar gas, dragging the
magnetic field with the gas are good candidates for injection of
turbulence into the GMF on scales of about 100~pc. Such structures
exist throughout the entire Galaxy and also affect the large-scale
structure of the magnetic field. Further candidates for injecting
turbulence on larger scales are density wave shocks \cite{kim06} and 
gravitational instability in rotating
discs \cite{larson85,lin87,lesch93}. We estimate typical length
scales of the latter in our Galaxy, which compare to our maximum length
$L_\mathrm{max}=1~\mathrm{kpc}$ of the small-scale field. 
In addition
to the recent progress in modelling the GMF, numerical simulations of
galaxies are starting to contribute to our understanding of the
seeding of magnetic fields in collapsed cosmic structures (see
e.g. \cite{beck12,beck13a,pakmor14}) and will hopefully soon provide us with
reasonable models of magnetic fields in galaxies.

Our resulting length scales of the (an)isotropic random field and the
recent work of \cite{mertsch13} strengthen the argument that
supernovae influence features in synchrotron and RM full-sky maps. In
recent studies large coherence lengths are favoured and play an
important role in the analysis of the small-scale
GMF. They range from 20 -- 100~pc \cite{keivani15} to 220~pc (our
study) and up to 420~pc \cite{mertsch13}. 
However, \cite{haverkorn04,haverkorn06,haverkorn08} find outer turbulent scales in 
spiral arms that seem to be much smaller than interarm regions, i.e. 
only a few parsec.
We also calculate the mean distance $L1$ of the RM signal power 
spectrum curves in units of $\sigma$ (see Fig. \ref{fig:spec}) and obtain 
the following values $L1$= 17.0, 5.15, 2.24, 0.915, and 0.609 for $L_{c}$ 
values ranging from 20 pc to 500 pc and, regarding the RM signal power 
spectrum, state that $\approx$ 220~pc is an lower 1-$\sigma$ limit for $L_{c}$, 
whereas confidence levels for 100~pc and 50~pc are about 2 and 5 $\sigma$.
Much larger scales would lead to RM values above the galactic plane, which significantly 
exceed the observed ones (Fig. \ref{fig:rm_lc}). 
A similar trend can be found in polarized emission maps, with best matching scales 
of $\approx$ 220~pc (I 408 MHz) and $\approx$ 100~pc (PI WMAP) (Fig. \ref{fig:pi_spectrum}).

Additionally, detailed models of magnetic fields are important for studying
the sources and propagation of Galactic and extragalactic CR. Detailed
knowledge is also necessary to predict the distributions of arrival
directions of cosmic rays of energies above $10^{19}$ eV.
Astronomy with charged particles is possible if large deflection
angles cover only an insignificant fraction of the sky which would allow us
to trace back the charged particles due to relatively small (or well known)
deflections. In the last decade, a new generation of UHECR observatories
have come into operation: the Pierre Auger observatory in the Southern
hemisphere and the Telescope Array in the Northern hemisphere. In the
future it may be possible to constrain the GMF structure from direct
observations of arrival directions of UHECR with this new generation
of CR observatories (for recent results and future prospects see
e.g. \cite{letessier13}).

The use of the NE2001 model is still under debate. Besides the study of \cite{gaensler08} another study of \cite{savage09} suggests a modification of the scale height, namely 1.4 kpc, and a corresponding mid-plane density of about 0.016 cm$^{-3}$.
\cite{sun10} revisit the results of their 2008 GMF model with improvements to NE2001, i.e. the change of the scale height and the mid-plane density for the thick disk adapted from \cite{gaensler08}. 
As a consequence the reduced scale height of $n_\mathrm e$, the maximum hale field was reduced from 10 $\mu$G to 2 $\mu$G, which is considered by the authors to be more physically. 
Although they find improvement of the model using the adapted NE2001 model parameters an increase of the mid-plane thermal electron density by factor of about two is suggested to better reproduce the RMs in the Galactic plane.
The NE2001 model parameters are still under debate and we want to emphasize that the investigations of magnetic field models always depend on the uncertain thermal electron density models.

Furthermore there is progress in the investigation of the large-scale structure of the disk GMF. The very recent Planck results in \cite{jaffe16} give a systematic comparison of existing GMF models. In \cite{sun15} it is found that the field reversal at the equator (i.e. latitudes of about $\pm$ 30 degree) is an effect of the foreground.
They show that vertical fields do exist in our Galaxy like in nearly all observed external galaxies.
It is still under debate if antisymmetric fields like dipole fields are present. 
This means that otherwise the vertical field has to be quadrupole (which is what is predicted by dynamo theory for flat objects) or they come from Parker loops which lead to vertical field structures or from open filaments.
They also find that the JF12 large scale model fails to reproduce the large-scale halo field if polarization and Faraday rotation signatures of the North Polar Spur are subtracted. 
Therefore it is under debate, if the subtraction of the HI bubble of \cite{wolleben10} and the masking of pixels in the Stokes Q and U full sky maps which JF12 associates with the North Polar Spur is sufficient enough.

\section{Summary \& Conclusions}

We use the HAMMURABI code to compute full-sky maps of total, polarized
synchrotron intensity, rotation measure maps and cosmic ray deflection
maps. For this, spatial distributions of cosmic ray electrons, thermal
electrons and a model for the Galactic magnetic field (GMF) are required. Our
special focus here lies in the small-scale GMF and how to incorporate it 
into the existing magnetic field models for the Milky Way.
Previously to this work, the small-scale field was modelled as a scalar
field that resembles the RMS value of this field and a macroscopic
prescription was used which describes the influence of this small-scale
field model on the observables of total and polarized synchrotron
intensity. Following this formulation it was assumed that there
is no effect of (an)isotropic random fields on RM maps or isotropic 
random fields on polarized synchrotron maps. Therefore, no conclusions
about the length scales of the fluctuations in this fields nor an energy
spectrum can be drawn.  

We have introduced a new prescription for the modelling process which is on the
one hand capable to model the length scales of the random field and on the 
other hand allows a more explicit treatment of the influence of the small 
scale magnetic fields on predictions of key observables like total and
polarized synchrotron intensity and rotation measures. The computed maps
are compared to  observations and used to adjust some of the
key GMF model parameters. Additionally, our approach
allows  us to generate multiple realizations of the  small-scale field and
thereby allows to determine uncertainties to the key observables
in such maps. 

Comparing to the observed rotation measure distribution, our main
findings can be summarized as follows:

\begin{itemize}
\item Our best-fitting small-scale magnetic field model is characterized by
a correlation length of approximately 220~pc.  
\item The amplitude of the anisotropic magnetic field component in the 
JF12 model has to be scaled down by a factor $f_a\approx0.3$ while the 
amplitude of the isotropic magnetic field has to be scaled down by a 
factor $f_i\approx0.6$ to best compare to observations. 
\item With these parameters, our model produces for the first time
a rotation measure map which reproduces a power law power spectrum
which compares well with the observed one in slope and amplitude.
\item At the same time our new model for the first time reproduces
the latitude profile for the polarized synchrotron intensity and
produces small scale structures in the polarized intensity map which
are in qualitative agreement with the observations.
\item Our magnetic field model predicts a rotation measure of 
$-2.8\pm7.1$ rad/m$^2$ and $4.4\pm11.0$ rad/m$^2$ for the north
and south Galactic pole respectively, which is in good agreement
with recent observational findings.
\end{itemize}

The statistical properties of the GMF play an important role in
solving long-term astrophysical problems such as the propagation of
cosmic rays throughout our Galaxy. In this paper, we also show that
small-scale magnetic field components are not negligible. Our new
formulation of the random magnetic field 
allows us to additionally study the uncertainties of the deflection of UHECRs.
In particular we find that:
\begin{itemize}
\item The overall deflection of UHECRs  is slightly larger than
in the original JF12 model, and not adjusting the original model
when introducing a realization of the random field would yield 
larger general amplitudes of  UHECR deflections.
\item Calculating various realizations of the random magnetic field
component allows us to calculate uncertainties on the predicted deflection
map. For the region around CenA we inferred a mean deflection of 
$\approx3.5\pm1.1$ degree.  
\end{itemize}

\section{Outlook}

Thermal electron density fluctuations are expected to exist
and are commonly assumed to couple to magnetic field fluctuations 
(due to turbulent motions of the gas). Therefore in improving our 
predictions of the magnetic field fluctuations from such calculations we
would need to include the fluctuations within the electron density,
which are currently not included within the NE2001 description of the 
thermal electron distribution.
Although the spectrum of the fluctuations of the thermal electrons are
expected to follow a Kolmogorov-like power law \cite{armstrong95}, it
remains unclear how the magnetic field strength should be treated
across such different density regions. For example, \cite{sun09} 
apply in a similar approach the NE2001 thermal electron distribution
in combination with a Kolmogorov power spectrum for the turbulent
magnetic field. Such additional fluctuations in the
underlying electron density fluctuation are expected to leave additional
imprints within the observed values in the RM full-sky map. A detailed
discussion of the correlation between magnetic field strength and thermal
electron distribution can for example be found in \cite{beck03}. However
since turbulence in the thermal electron distribution is constrained by an
upper limit of about  30~pc \cite{armstrong95}, we conclude that larger
fluctuations of the magnetic field can be treated separately as in
our approach, and a more detailed model which includes fluctuation in the
electron density distribution can be left for future, more detailed
investigations.


\acknowledgments
We thank the anonymous referee for the comments, which helped to strengthen our argumentation and improve several sections of this paper.
We thank Glennys Farrar, Niels Oppermann, Ann Mao, Wolfgang Reich and Xiaohui Sun for useful comments and discussions.
AMB, KD and RB are supported by the DFG Research Unit 1254.
AMB and KD are supported by the DFG Cluster of Excellence `Origin and Structure of the Universe'.
The authors gratefully acknowledge the computing time granted by the John von Neumann Institute for Computing (NIC) and provided on the supercomputer JUROPA at J\"ulich Supercomputing Centre (JSC).

\appendix
\section{Conversion of cosmic ray electron data}
The cosmic ray electron distribution provided by GALPROP was converted to a form suitable for HAMMURABI as follows.
\begin{figure}
\begin{center}
  \includegraphics[width=0.475\textwidth]{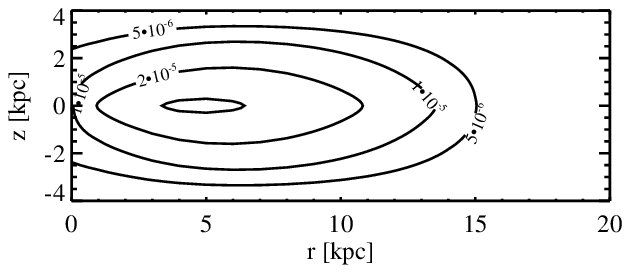}
  \includegraphics[width=0.475\textwidth]{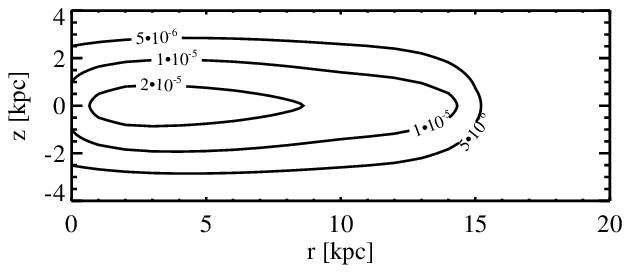}
  \caption{Spatial distribution of relativistic electrons in the Milky Way used by JF12 (left, courtesy of G. Farrar) and 
  calculated from particle spectra derived from the GALPROP model with ID 54\_z04LMPDS (right).
  The contour levels represent the density of electrons in terms of the coefficient $C$ in $N(\gamma)d\gamma = C(r) \gamma^{-p}$, units cm$^{-3}$.}
  \label{fig:ncre_galprop}
\end{center}
\end{figure}
HAMMURABI requires the density in the form $N(\gamma)d\gamma = C(r) \gamma^{-p}$, where $\gamma=E/m_e$ is the electron Lorentz factor, and $r$ is the position in the Galaxy.
Although the units are cm$^{-3}$, this quantity does not represent the actual density of CRe, but is a scaled representation of this.
GALPROP provides the flux $I(E)$ at $(R,z)$ in cm$^{-2}$ sr$^{-1}$ s$^{-1}$ MeV$^{-1}$ (the GALPROP FITS file has actually $E^2 I(E)$ in MeV$^2$ cm$^{-2}$ sr$^{-1}$ s$^{-1}$ MeV$^{-1}$).
Using  $N(\gamma)d\gamma=N(E)dE$, a power-law $E^{-p}$, and a factor $4\pi/c$ to convert from flux to density, we arrive at $C(r) = (4\pi/c) m_e  (E_o/m_e)^p I(R,z,E_0)$, where $E_0$ is a reference energy.
We use  $E_0=$ 10 GeV and $p=$ 3, as in HAMMURABI.
This is a reasonable approximation above 10 GeV, which is the principal range contributing to the synchrotron emission studied in this paper.
The input GALPROP file and HAMMURABI format for our model are available on request.
Fig. \ref{fig:ncre_galprop} shows the spatial distribution of $n_\mathrm{CRe}$.
We find a difference of factor 1.4 between the $n_\mathrm{CRe}$ data used by JF12 and our data.

\section{Synchrotron map power spectra}
The main constraint for UHECR deflection comes from RMs, therefore RM full sky maps and power spectra are investigated. In addition  we analyse the influence of the chosen magnetic field spectrum on the signal spectrum of the PI map.
\begin{figure}
\begin{center}
  \includegraphics[width=0.475\textwidth]{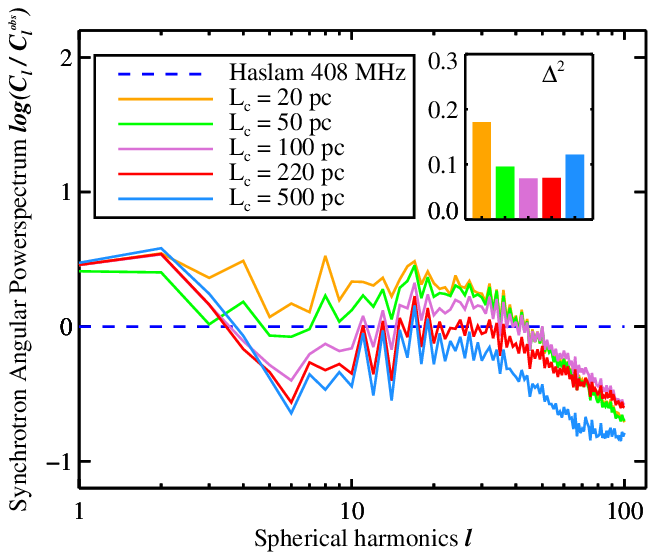}
  \includegraphics[width=0.475\textwidth]{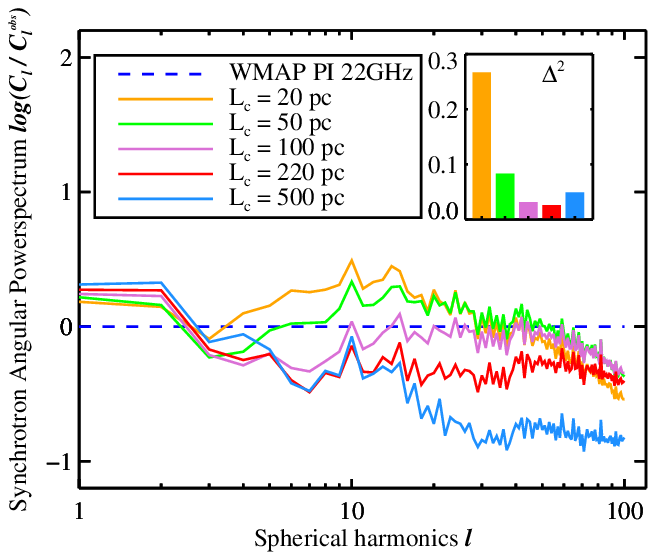}
  \caption{Mean ensemble power spectra of the synthetic total I signal maps (left) and PI signal maps (right) for various length scales of the turbulent magnetic field power spectrum. 
  We normalise all the mean power spectra to values derived from the cleaned Haslam 408 MHz total intensity \cite{remazailles14} respectively the WMAP 22~GHz polarized synchrotron emission map.}
  \label{fig:pi_spectrum}
\end{center}
\end{figure}
Fig. \ref{fig:pi_spectrum} shows the obtained power spectra for various length scales of the small scale magnetic field.
Thereby the maps are normalized by their own latitude profile (see Fig. \ref{fig:profiles_synchrotron}) in order to take account for the differing total strength of the signal in the map. 
The power spectra are then compared to observation, i.e. the total intensity spectra are compared to the spectrum obtained from Haslam 408~MHz data \cite{remazailles14} respectively the WMAP 22~GHz dataset.
The residual spectra, i.e. difference between simulated mock spectrum and observational spectrum, are shown. 
That means, that the slope of the simulated spectral curves coincide if they are horizontal in the residual plot.
We find that our best-fitting correlation length of 220~pc produces the best horizontal residual spectrum for 22~GHz although the curve is slightly shifted downwards for $l \gtrsim 3$. For 408~MHz we find well fitting for $l$ between about 10 to 50 and a lack of power on greater and lower angular scales.
\cite{mertsch13} also studied fluctuations in synchrotron maps and added supernova shells and turbulent Kolmogorov fields to obtain the spectrum of the uncleaned 408~MHz map of \cite{haslam81} on small and very small angular scales.


  \bibliography{paper}
  \bibliographystyle{JHEP}

\label{lastpage}

\end{document}